\documentclass[a4paper,11pt]{article}
\usepackage{jheppub,verbatim,mathrsfs}

\title{Anomalous Breaking of Anisotropic Scaling Symmetry in the Quantum Lifshitz Model}

\author{Marco Baggio,}
\author{Jan de Boer}
\author{and Kristian Holsheimer}

\affiliation{Institute for Theoretical Physics, University of Amsterdam,\\
Science Park 904, Postbus 94485, 1090 GL Amsterdam, The Netherlands}

\emailAdd{m.baggio@uva.nl}
\emailAdd{j.deboer@uva.nl}
\emailAdd{k.holsheimer@uva.nl}

\abstract{
In this note we investigate the anomalous breaking of anisotropic scaling symmetry $(t,x)\rightarrow(\lambda^z\,t,\lambda\,x)$ in a non-relativistic field theory with dynamical exponent $z=2$. On general grounds, one can show that there exist two possible ``central charges'' which characterize the breaking of scale invariance. Using heat kernel methods, we compute these two central charges in the quantum Lifshitz model, a free field theory which is second order in time and fourth order in spatial derivatives. We find that one of the two central charges vanishes. Interestingly, this is also true for strongly coupled non-relativistic field theories with a geometric dual described by a metric and a massive vector field. 
}

\arxivnumber{1112.6416}

\newcommand{\bea}{\begin{eqnarray}}
\newcommand{\eea}{\end{eqnarray}}
\newcommand{\be}{\begin{equation}}
\newcommand{\ee}{\end{equation}}

\begin{document}

\maketitle


\section{Introduction}
In recent years there has been considerable progress in extending the holographic correspondence to spacetimes that are not asymptotically AdS. Schroedinger \cite{Son:2008ye,Balasubramanian:2008dm} and Lifshitz \cite{KLM,Koroteev:2007yp} spacetimes were introduced as possible duals of non-relativistic strongly coupled field theories.

Holographic renormalization for these spacetimes was explored in \cite{Guica:2010sw} for the Schr\"o\-ding\-er case and more recently in \cite{Ross:2011gu,Baggio:2011cp,Mann:2011hg} for the Lifshitz case. In particular, it was suggested in \cite{Baggio:2011cp} that the non-relativistic analog of the Weyl anomaly, derived in the context of ordinary AdS/CFT in \cite{Henningson:1998gx}, might be present for 3+1 dimensional Lifshitz spacetimes with $z=2$.
Conformal anomalies play an important role in relativistic field theories, especially in two dimensions, where various physical quantities display
universal behavior which is governed by the central charge only. It is clearly of interest to explore to what extent similar results carry over
to non-relativistic field theories. 

Motivated by these observations, in this paper we explore the general structure of the Weyl anomaly for non-relativistic field theories in $d=2+1$
dimensions with $z=2$. The cases $d=4$, $z=3$ and $d=6$ were analyzed in \cite{Adam:2009gq} and \cite{Gomes:2011di} respectively. Contrary to the relativistic case, where anomalies are present only for even dimension, we show that in the non-relativistic setting anomalies can be generated in odd dimension as well.
That this is in principle possible can be quite easily seen by dimensional analysis. The analogue of the trace of the stress-tensor for non-relativistic
field theories has dimension $z+d-1$. A term with $a$ time derivatives and $b$ spatial derivatives, on the other hand, has dimension $az+b$. For 
generic $z$, there can only be contributions to the conformal anomaly with $a=1$ and $b=d-1$. Such terms can only appear in theories which
explicitly break time reversal invariance, which will not be the case for the theories that we consider. For special values of $z$, especially for
integer values of $z$, other values of $a,b$ are allowed: for example, for $d=z+1$, terms with either $(a,b)=(2,0)$ or $(a,b)=(0,2z)$
can appear. In particular, for the case of interest to us with $z=2$, $d=3$, one can have terms with either two time or four spatial derivatives.
There are many such terms that one can write down, but the more detailed analysis that we give below shows that 
the anomaly is generated by a total of two linearly independent and non-trivial structures.

In order to show that the anomaly is indeed generated in field theory models enjoying Lifshitz symmetry, we compute the heat-kernel expansion of a fourth-order differential operator that transforms covariantly under local anisotropic rescalings. The corresponding free field theory model has various applications in condensed matter physics, see \cite{Ardonne:2003wa} and references therein. We explicitly compute the numerical coefficients that appear in front
of the two possible contributions to the anomaly, and find that only one of them  is nonzero. 

An obvious question is then whether a similar statement holds for those
purported strongly coupled field theories which are dual to Lifshitz spacetimes. We compute the two central charges using the holographic renormalization methods of \cite{Baggio:2011cp}.
Just like in the free field theory example, only one of the coefficients turns out to be non-vanishing.
The results for the holographic computation perfectly agree with those of the recent paper \cite{Griffin:2011xs}, which computes the full conformal anomaly using the methods of \cite{Ross:2011gu} and which appeared while we were preparing this paper for
submission to the ArXiv. 

Remarkably, the anomaly in both models is entirely generated by terms that only contain time derivatives, even though more general structures containing space derivatives are in principle allowed. It is not clear to us whether there is a simple explanation of this fact.

The outline of this paper is as follows. In section~2, we setup and perform the field theory computation of the conformal anomaly. The corresponding
analysis for Lifshitz spacetimes is presented in section~3, and in section~4 we present our conclusion. Some technical details are discussed in the
appendices, such as the complete classification of the possible structures that can appear in the conformal anomaly, and an interpretation of the
bulk metric and gauge field in the field theory language. 

\textbf{Note added:} In an earlier version of this paper, we erroneously claimed that there were three instead of two possible conformal anomalies and used holographic renormalization to compute two of the three central charges. The aforementioned paper \cite{Griffin:2011xs} correctly pointed out that there can only be two and not three contributions to the anomaly. We corrected this mistake in this version and as a result our original anomaly computation now yields the complete anomaly.



\section{Anisotropic-Scaling Anomaly in the Quantum Lifshitz Model}
\label{sec:heatkernel}
Scale invariance arises naturally when describing systems at a quantum critical point. Even when the underlying theory is \emph{not} Lorentz invariant, it may still exhibit invariance under so-called Lifshitz-like scaling that treats time and space differently,
\begin{align}\label{eq:lifshitz-scaling}
\vec{x}&\mapsto \lambda\,\vec{x}, &
t\mapsto\lambda^z t.
\end{align}
Only the specific value $z=1$ is compatible with Lorentz symmetry. In the high-energy-physics community \eqref{eq:lifshitz-scaling} is often called anisotropic scaling, though in the consensed-matter literature the word `anisotropy' is typically used for \emph{spatial} anisotropy only. 

In this section we analyze the anomalous breaking of anisotropic scaling symmetry in a three-dimensional field theory known as the \emph{quantum Lifshitz model} \cite{Ardonne:2003wa}. This model arises in the classical field theory description of three-dimensional Lifshitz points, which is why the corresponding quantized theory is called the `quantum Lifshitz model'. We refer the interested reader to \cite{Ardonne:2003wa} and references therein for a discussion on applications of the quantum Lifshitz model to condensed matter systems such as the so-called quantum dimer model.

We shall find that the anomalous breaking of the Lifshitz-like scaling \eqref{eq:lifshitz-scaling}, which we will sometimes refer to as the `Lifshitz anomaly', is parametrized by two central charges. In this section we will use heat kernel methods in order to determine these two central charges in the quantum Lifshitz model.

\subsection{The Quantum Lifshitz model}
We consider a three dimensional free boson $\phi(t,x)$ with the following Euclidean action
\be
\label{eqn:lifshitzmodel}
S=\int dt d^{2}x N\sqrt{h} \left(\frac{1}{2} N^{-2}(\partial_{t} \phi)^{2} + \frac{1}{2} (\Delta \phi)^{2}\right),
\ee
where we have coupled the system to the degrees of freedom of an auxiliary 
three-dimensional metric with a preferred time foliation, of the form:\footnote{We did not include a shift $N^i$ , because it can be removed locally by a foliation preserving diffeomorphism, e.g. $t\mapsto \tau(t)$ and $x^i\mapsto \xi^i(t,\vec{x})$.}
\be \label{eqn:auxmetric}
ds^{2} = N^{2} dt^{2} + h_{i j} dx^{i} dx^{j}.
\ee
Furthermore, $\Delta = \nabla_{i}\nabla^{i}$ is constructed out of the covariant derivatives of the spatial metric $h_{ij}$. 
The metric \eqref{eqn:auxmetric} keeps its form under diffeomorphisms in time, $t\mapsto \tau(t)$, and in space, $x^i \mapsto \xi^i(\vec{x})$, and
both leave the action (\ref{eqn:lifshitzmodel}) invariant if we transform $N$ and $h_{ij}$ as dictated by \eqref{eq:lifshitz-scaling}. This is one of the reasons
why it is convenient to combine $N$ and $h_{ij}$ in an auxiliary metric, the other reason is that this structure also naturally emerges in the
holographic setup.

By integrating by parts and ignoring boundary terms\footnote{In the following, we will assume that the theory is defined on a manifold without boundary.}, the action can be written as:
\be
\label{eqn:quadraticform}
S=\int dt d^{2}x\, N\sqrt{h}\, \phi \, D \, \phi,
\ee
where $D$ is given by:
\be
\label{eqn:diffoperator}
D = -\frac{1}{N \sqrt{h}} \partial_{t} N^{-1}\sqrt{h}\, \partial_{t} + \frac{1}{N} \Delta N \Delta
\ee
This model is classically invariant under local anisotropic scale transformations, that is
\begin{align}
N & \to e^{2 \omega} N, &
h_{ij} & \to e^{2 \omega} h_{ij},&
\phi & \to \phi,
\end{align}
where $\omega$ is an arbitrary function of $t$ and $x^{i}$. In particular, the operator $D$ transforms as
\be
D \to e^{-4\omega} D
\ee
The variation of the action $S$ under a Weyl transformation is thus given by
\be
\delta S = \int dt d^{2}x \left(\delta N \frac{\delta S}{\delta N} + \delta h_{ij} \frac{\delta S}{\delta h_{ij}}\right) = \int dt d^{2}x\, \delta \rho \left(2 N \frac{\delta S}{\delta N} + 2 h_{ij} \frac{\delta S}{\delta h_{ij}} \right).
\ee
Defining the energy density $\mathcal{E}$ and momentum flux (spatial stress tensor) $\Pi_{ij}$ as
\begin{align}\label{eq:def-energy-stress}
\mathcal{E}\ & = \frac{2}{N\sqrt{h}}\; N^{2} \frac{\delta S}{\delta N^{2}}, &
\Pi^{ij} & = \frac{2}{N\sqrt{h}}\; \frac{\delta S}{\delta h_{ij}},
\end{align}
we see that local anisotropic scale invariance implies:
\be
2\mathcal{E} + \Pi^{i}_{i} = 0,
\ee
which is the non-relativistic analog of the tracelessness condition $T^{a}{}_{a} = 0$.

We will prove in this section that the classical anisotropic conformal invariance of this model is broken at the quantum level. The quantum expectation values of the energy density $\left<\mathcal{E}\right>$ and the spatial stress tensor $\left<\Pi^{ij}\right>$ are given by \eqref{eq:def-energy-stress} when one replaces $S$ by the effective action $W=-\log Z$. In the presence of an anomaly, the right-hand side of
\be
\label{eqn:traceanomaly}
2\left<\mathcal{E}\right> + \left<\Pi^{i}_{i}\right> = \mathcal{A}
\ee
is non-zero, where $\mathcal{A}$ is the anomaly. Equivalently, one may express the anomaly directly in terms of the variation of the (renormalized) effective action $W$,
\be
\delta W = \int N \sqrt{h}\, \delta\rho\, \mathcal A,
\ee
where $\delta \rho = \omega$ is the infinitesimal scale factor. We will show that
\be
\mathcal{A} = \frac{1}{1536\pi} \frac1{N^2}\left(16 h^{ij}N\partial_t(N^{-1}\dot h_{ij}) + 5(h^{ij} \dot{h}_{ij})^{2} - 10 h^{ij}\dot{h}_{jk} h^{kl}\dot{h}_{li}\right) + \frac{1}{480\pi} \frac{1}{N}\nabla_{i} J^{i},
\ee
where a dot indicates $\partial_{t}$ and $\nabla_{i} J^{i}$ is a ``trivial'' total derivative, by which we mean that it can be removed by adding appropriate local counterterms. In fact we will show that by adding appropriate counterterms to the action \eqref{eqn:lifshitzmodel}, the anomaly can be written in the simpler form
\be
\mathcal{A} = \frac{1}{128\pi} \frac{1}{N^{2}}\left(h^{ij}h^{kl}\,\dot{h}_{ik}\dot{h}_{jl} - \frac12(h^{ij} \dot{h}_{ij})^{2} \right).
\ee 
This anomaly will be computed using a heat-kernel expansion.

As an aside, one of the reasons for the particular interest in this model is that the ground-state wave functional is invariant under time-independent conformal transformations in space. All equal-time correlators can be computed using the machinery of a two-dimensional field theory \cite{PhysRevB.23.4615,Ardonne:2003wa}. One may thus naively expect that the anomalous breaking of anisotropic scaling symmetry \eqref{eq:lifshitz-scaling} is somehow related to the two-dimensional Weyl anomaly $\langle T^i{}_i\rangle\propto R$, see e.g.\ \S 5.A of \cite{DiFrancesco:1997nk}. We find, however, that this is \emph{not} the case. The anomaly that we find involves only derivatives with respect to the \emph{time} coordinate, whereas the two-dimensional Ricci scalar $R$ obviously only contains spatial derivatives.

\subsection{Heat-kernel expansion}
The quantum effective action $W$ for the model \eqref{eqn:lifshitzmodel} can be computed explicitly, and is  given by the formal expression:
\be
W \propto \frac{1}{2} \ln {\rm det} (D),
\ee
where ${\rm det} (D)$ is the determinant of the operator $D$ defined in equation \eqref{eqn:diffoperator}. As usual, this determinant is not well-defined and must be regularized. We will employ $\zeta$-function regularization. We define the generalized zeta function as
\be
\zeta(s,f,D) = {\rm Tr}_{L^{2}} (f D^{-s}),
\ee
where $s$ is an arbitrary positive number and $L^{2}$ an appropriate function space on which $D^{-s}$ is trace-class. The regularized effective action is given by \cite{Vassilevich:2003xt}:
\be
W = -\frac{1}{2}\zeta'(0,1,D) - \frac{1}{2}\ln(\mu^{2}) \zeta(0,1,D),
\ee
where $\zeta'(0,f,D)=\partial_{s}\zeta(s,f,D)|_{s=0}$ and $\mu$ is the usual arbitrary renormalization scale. The zeta function $\zeta(s,f,D)$ is related via a Mellin transformation to the heat kernel:\footnote{In particular, the relation between the two is
\begin{align*}
\zeta(s,f,D)\ &=\ \Gamma(s)^{-1}\int_0^\infty d\epsilon\,\epsilon^{s-1}\,K(\epsilon,f,D),&
K(\epsilon,f,D)\ &=\ \frac1{2\pi i}\oint ds\,\epsilon^{-s}\,\Gamma(s)\,\zeta(s,f,D).
\end{align*}
}
\be
K(\epsilon,f,D)= {\rm Tr}_{L^{2}}(f\, e^{-\epsilon D}),
\ee
where $f$ is an arbitrary function of $t$ and $x^{i}$, and $\epsilon$ is an arbitrary positive parameter. In principle $K$ depends on the global behavior of the operator $D$ (the trace can be written as a sum over the spectrum of the operator, which is determined by global properties); however there is an asymptotic series of the form:
\be
K(\epsilon,f,D) \sim \sum_{k=0}^{\infty} \epsilon^{\frac{k}{2}-1} \tilde a_{k}(f,D),
\ee
where $\tilde a_{k}(f,D)$ can be computed \emph{locally} from $N$ and $h_{ij}$. By repeating the analysis of \cite{Vassilevich:2003xt} section 7.1, it is possible to show that the variation of the renormalized effective action under an infinitesimal anisotropic local scale transformation $h \to (1+ 2 \delta\rho)h$, $N \to (1+2\delta \rho)N$, is given by \footnote{The factor 2 comes from the factor 4 in $D \to e^{-4\rho} D$ under scale transformations.}:
\be
\label{eqn:anomtrans}
\delta W = - 2 \tilde a_{2}(\delta\rho, D).
\ee
In other words, the anomaly is given by the $\epsilon^{0}$ term in the heat-kernel expansion. As explained above, this will be a local functional of $N$ and $h$; we will therefore write:
\be
\tilde a_{2}(f,D) = \int dt d^{2}x N\sqrt{h} f\, a_{2}(N,h_{ij}),
\ee
where $a_{2}(N,h_{ij})$ is a local function that depends on $N$ and $h_{ij}$. The Weyl variation of the marginal heat kernel coefficient $\tilde{a}_2(1,D)$ vanishes. To see why, consider the Weyl variation of the heat kernel coefficient at order $k$,
\begin{align}
\frac{d}{d\gamma}\bigg|_{\gamma=0}\tilde{a}_k(1,e^{-4\gamma f}D)\ =\ (4-2k)\,\tilde{a}_k(f,D),
\end{align}
which vanishes when $k=2$. We used the identity for the Weyl variation of the full heat kernel,
\begin{align}
\frac{d}{d\gamma}\bigg|_{\gamma=0}K(\epsilon,1,e^{-4\gamma f}D)
&=\ \frac{d}{d\gamma}\bigg|_{\gamma=0}\text{Tr}\!\left( e^{-\epsilon\,\exp(-4\gamma f)D} \right)\\ 
&=\ 4\epsilon\,\text{Tr}\!\left( f D\, e^{-\epsilon D} \right) \\ 
&=\ -4\epsilon\frac{d}{d\epsilon}\,\text{Tr}\!\left(f\,e^{-\epsilon D}\right) \\
&=\ -4\epsilon\frac{d}{d\epsilon}K(\epsilon,f,D).
\end{align}
As a consequence, we can use this property that $\delta \tilde a_{2} (1,D) = 0$ to find that, in local language,
\be
\label{eqn:localweyl}
N \sqrt{h}\, a_{2}(N,h_{ij})\ =\  e^{4 \omega} N\sqrt{h}\, a_{2}(e^{2\omega}N,  e^{2\omega}h_{ij})  + \text{total derivatives}.
\ee
In other words, the anomaly itself must be Weyl invariant (up to total derivatives), and this puts very strong restrictions on the possible terms that can appear in $a_{2}$.

\subsection{Analysis of the possible terms}
Just by dimensional analysis (i.e. by requiring invariance under constant rescalings) we can see that there are many terms
of the right dimension that can appear in the heat kernel
expansion. In dimensional analysis, $\partial_t$ has dimension
two and $\partial_i$ has dimension one; we are interested
in terms of dimension four that are covariant under reparametrizations
of $t$ and reparametrizations of $x^{i}$. These reparametrizations should
not mix $t$ and $x^i$ since that would ruin the form of $D$. 
Examples of terms of the right dimension are\footnote{A complete classification can be found in Appendix \ref{sec:partialintegration}.}
\begin{equation}
\begin{array}{l}
N^{-2} \partial_{t} h_{ij} G^{ijkl} \partial_{t} h_{kl},\\[10pt]
R^2,\qquad N^{-1}\Delta N\,R,\qquad (N^{-1}\Delta N)^2,\\[10pt]
h^{ij} h^{kl} (N^{-1} \partial_i N) (N^{-1} \partial_k N)(N^{-1} \nabla_j\partial_l N).\\[10pt]
\ldots
\end{array}
\ee 
where $G^{ijkl} = \frac{1}{2} (g^{ik} g^{jl} + g^{il} g^{jk})-\lambda g^{ij} g^{kl}$ is the DeWitt metric and $\lambda$ an arbitrary real number (in General Relativity $\lambda = \frac{1}{D-2}$, where $D$ is the spacetime dimension).
Terms with one time derivative and two space derivatives cannot appear because of time reversal symmetry. All these terms are scale invariant, but the anomaly should also obey the Wess-Zumino consistency condition (namely the second variation of the effective action $W$ should be symmetric). The possible consistent structures are classified in Appendix \ref{sec:partialintegration}. The most general form of the anomaly is therefore:
\begin{align}
\label{eqn:genanomaly}
\mathcal{A}\ &=\ C_{1}\, \frac{1}{N^{2}}\left(h^{ij}h^{kl}\,\dot{h}_{ik}\dot{h}_{jl}-\frac12(h^{ij} \dot{h}_{ij})^{2} \right) \nonumber\\
&\ +  C_{2}\, \left(R - \frac{1}{N} \Delta N + h^{ij}\left(\frac{1}{N} \partial_{i} N\right) \left(\frac{1}{N} \partial_{j} N\right)\right)^{2} \\
&\ + (\text{trivial total derivatives}).\nonumber
\end{align}
Here the trivial total derivatives can be cancelled by appropriate counterterms.


\subsection{Computation of the anomaly}

As we explained above, the spectral function can be expanded as:
\be
K(\epsilon,f,D)=\sum_{k \geq 0} \epsilon^{\frac{k}{2}-1} \int dt d^2 x N\sqrt{h} f(t,x) a_{k}(N,h_{ij}),
\ee
where $a_{k}(N,h_{ij})$ is a local function of $N$ and $h_{ij}$. To evaluate this we need a suitable basis; it is customary to use the rescaled Fourier modes so that they are orthonormal with respect to the measure that includes
the $N\sqrt{h}$ factor. Nevertheless, as pointed out in \cite{Ceresole:1988hn}, the cyclicity of the trace allows us to use the usual flat Fourier modes. We thus find
\be
K=\int\frac{d\omega d^2 k}{(2\pi)^3} \int dt d^2 x\,
e^{-i\omega t-ikx} f e^{-\epsilon D} 
e^{i\omega t+ikx}.
\ee
We can conjugate the Fourier mode to the left to get the expression
\be
K=\int\frac{d\omega d^2 k}{(2\pi)^3} \int dt d^2 x\,
f e^{-\epsilon D_2},
\ee
where $D_{2}$ is obtained from $D$ by shifting the derivatives as follows:
\begin{align}
\partial_{t} & \to \partial_{t} + i \omega &
\partial_{i} & \to \partial_{i} + i k_{i}.
\end{align}
The most singular term in the heat kernel is the one where we keep only the terms in $D_2$ without derivatives, leading to
\be
\frac{1}{\epsilon}\tilde{a}_0(f,D) = \int\frac{d\omega d^2 k}{(2\pi)^3} \int dt d^2 x f e^{-\epsilon(N^{-2}\omega^2 +(k^2)^2)},
\ee
where $k^2\equiv h^{ij}k_i k_j$. This expression is readily evaluated to yield the first term in the heat kernel expansion:
\be
\tilde{a}_0(f,D) = \frac{1}{16\pi} \int dt d^2 x N \sqrt{h} f(t,x)
\ee
Computing the subleading terms in the heat
kernel expansion is now straightforward though
somewhat involved. We shall write 
\be
D_2 = D_2^0 + D_2^\text{int}
\ee 
where $D_2^0$ is the piece we isolated above
that contains $\omega^2$ and $k^4$, and $D_2^\text{int}$
the remainder. We then expand the exponential
of $D_2^\text{int}$. It contains a factor of $\epsilon$,
but $\omega$ counts as $\epsilon^{-1/2}$ and $k$ as
$\epsilon^{-1/4}$ in the Gaussian integral, therefore
$D_2^\text{int}$ has a term which scales as
$\epsilon^{-1/4}$, and to get to the finite term
one needs to expand $D_2^\text{int}$ up to fourth 
order so that we get terms up to $k^{12}$.
However, the problem becomes tractable if we consider the time-derivative and space-derivative sectors separately. This is consistent because the anomaly can only have structures involving either \emph{two} time derivatives or \emph{four} spatial derivatives.

\subsubsection*{The two-derivative anomaly}
In order to compute the two-derivative contribution to the anomaly, and in turn $C_{1}$, it is sufficient to consider metrics that only
depend on $t$, and not on $x^a$. Thus
we can drop all the terms with spatial derivatives $\partial_{i}$ in $D_{2}^\text{int}$. Moreover,
by changing the coordinate $t$ if
necessary, we can take $N=1$. 
With these assumptions, we have: 
\begin{align}
D_2^{0} & =\omega^2 + (k^2)^2,\\
D_2^\text{int} & =-i\omega(\partial_t + \frac{1}{\sqrt{h}} \partial_t \sqrt{h} ) - \frac{1}{\sqrt{h}} \partial_t \sqrt{h}\partial_t.
\end{align}
We have to expand to second order in $D_2^\text{int}$. Since $D_2^\text{int}$ and $D_2^0$ do not commute, we use the following formula:
\be
e^{A+B} = e^A + \int_{0\leq \alpha\leq 1} d\alpha\,
e^{\alpha A} B\, e^{(1-\alpha)A} + 
\int_{0\leq \alpha+\beta\leq 1} d\alpha\, d\beta\, e^{\alpha A} B\, e^{\beta A}
B\, e^{(1-\alpha-\beta)A} + {\cal O}(B^3).
\ee
We find the following contribution to $a_{2}$:\footnote{This computation is in principle quite lengthy. However, since there are only few terms that can appear, one can work this out for a diagonal $h_{ij}$ and then reconstruct the full answer.}
\be \label{aux1}
\tilde{a}_2(f,D)\ =\ \frac{-1}{1536\pi} \int dt d^2 x \sqrt{h} f \left\{ 
16 h^{ij} \ddot{h}_{ij} + 5 (h^{ij}\dot{h}_{ij})^2 -
10 h^{ij} \dot{h}_{jk} h^{kl} \dot{h}_{li} \right\} +\ldots
\ee
where the ellipses denote possible four-derivative contributions. To reinstate $N$, we simply need to change $dt \rightarrow
dt N$ and $\partial_t \rightarrow N^{-1} \partial_t$.
We can remove the first term by adding local counterterms as explained in Appendix \ref{sec:partialintegration}, to obtain the two-derivative contribution to the anomaly:\footnote{There is a relative factor -2, see \eqref{eqn:anomtrans}.}
\be\label{aux1-after-part-intgr}
\frac{1}{128\pi} \frac{1}{N^{2}}\left(h^{ij}h^{kl}\,\dot{h}_{ik}\dot{h}_{jl} - \frac12(h^{ij} \dot{h}_{ij})^{2} \right).
\ee
Using \eqref{eqn:anomtrans} and \eqref{eqn:genanomaly}, we see that
\be
C_{1} = \frac{1}{128 \pi}.
\ee

\subsubsection*{The four-derivative anomaly}
We now determine the four derivative contribution, and in turn $C_{2}$. As explained in Appendix \ref{sec:partialintegration}, there are 6 possible terms that can appear, 5 of which are total derivatives. These structures are distinguished by a metric of the form $h_{ij} = e^{2 f(x)} \delta_{ij}$ and $N = e^{g(x)}$, which can be used to simplify considerably the computation. The four-derivative contribution to the anomaly is thus
\be
\mathcal{A} = \frac{1}{480\pi} \frac{1}{N}\nabla_{i} \left(-5 (\partial^{i}N) R + 3 (\partial^{i} N) (\frac{1}{N} \Delta N) + 2 (\partial^{j} N) (\frac{1}{N}\nabla_{j}\partial^{i} N) - 5 \partial^{i} \Delta N\right).
\ee
It is interesting to note that this result is a total derivative, and as predicted by the Wess-Zumino consistency condition, it is orthogonal\footnote{To see what we mean by `orthogonal', we refer to the appendix \ref{sec:partialintegration}.} to the non-trivial total derivative $\mathcal{J}$ defined in equation \eqref{eqn:nontrivialcurrent}. As a consequence, this term can be removed by a local counterterm and we conclude that
\be
C_{2} = 0.
\ee
In Appendix \ref{sec:alternative} we present an alternative derivation of $C_{2}=0$.

\subsubsection*{\emph{The} anomaly}
In summary, the Lifshitz model \eqref{eqn:lifshitzmodel} exhibits an anomaly under anisotropic local scale transformations, which after the addition of appropriate counterterms is given by:
\be
\mathcal{A} = \frac{1}{128\pi} \frac{1}{N^{2}}\left(h^{ij}h^{kl}\,\dot{h}_{ik}\dot{h}_{jl} - \frac12(h^{ij} \dot{h}_{ij})^{2} \right).
\ee
It is striking that the anomaly involves only time derivatives. So far, it is unclear to us why this happens. It is also in contrast to the naive expectation that the anomaly is somehow related to the trace anomaly of a two-dimensional conformal field theory, as we mentioned at the beginning of this section.



\section{Holographic Calculation}
In the previous section we showed that a theory with anisotropic scaling symmetry has an anisotropic scaling symmetry anomaly parametrized by two central charges, denoted by $C_{1}$ and $C_{2}$. We computed these central charges for a particular model defined by the action \eqref{eqn:lifshitzmodel}. In this section we show that these central charges can be computed holographically for the Lifshitz spacetime considered in \cite{Ross:2011gu,Baggio:2011cp}.

\subsection{The Hamilton--Jacobi Formalism}
Lifshitz spacetime is a proposed gravitational dual to a field theory at a UV fixed point with anisotropic (Lifshitz-like) scaling symmetry $t\mapsto\lambda^zt$, $\vec{x}\mapsto\lambda\,\vec{x}$. The configuration of $(d+1)$-dimensional Lifshitz spacetime that we consider is a solution of the field equations derived from the Einstein--Proca action,
\begin{align}\label{eq:einstein-proca-action}
S\ =\ \frac{1}{16\pi G}\int d^{d+1}x\sqrt{-g}\left( R-2\Lambda -\frac14F_{\mu\nu}F^{\mu\nu}-\frac{m^2}2 A_\mu A^\mu \right)
\end{align}
The parameters of the theory are the curvature scale $\ell$ and the dynamical exponent $z$; they are related to the cosmological constant and the Proca mass via
\begin{align}
\Lambda\ &=\ -\frac{z^2+(d-2)z+(d-1)^2}{2\ell^2}, &
m^2\ &=\ \frac{(d-1)z}{\ell^2}.
\end{align}
We shall work in units such that $16\pi G=1$ and $\ell=1$ henceforth. Also, the computation that we shall perform in this section shall be for the case of $d=3$ boundary dimensions, though we would like to keep $d$ explicit where this may be illuminating. The solution is then given by
\begin{equation}\label{eq:lifshitz-solution}
ds^2\ =\ dr^2-e^{2zr}dt^2+e^{2r} d\vec x{\,}^2,\qquad\qquad A=\sqrt{-\alpha_0}\,e^{zr} dt.
\end{equation}
with $\alpha_0\equiv -(z-1)/z$. A shift in the radial coordinate $r\mapsto r+\log\lambda$ generates the Lifshitz scaling $t\rightarrow\lambda^zt$, $\vec{x}\rightarrow\lambda\vec{x}$.

The goal of this section is to compute the (divergent piece of the) on-shell value of the above action. We shall do this in the Hamilton--Jacobi (HJ) formalism \cite{deBoer:1999xf,Martelli:2002sp}. Much of this section should be seen as an extension to \cite{Baggio:2011cp}, so we refer to that work for a more detailed discussion of the HJ formalism. For instance, the ADM-like Hamiltonian associated to \eqref{eq:einstein-proca-action} was derived in \cite{Baggio:2011cp}. The HJ equation is a differential equation for the on-shell action, i.e.\ solving the HJ equation will give us the on-shell action $S_\text{cl}$. We write the HJ equation $H=0$ as\footnote{Again, see \cite{Baggio:2011cp} for more details. In particular, \eqref{eq:hj-eq-unexpanded} is the Hamiltonian constraint; the momentum constraint is automatically satisfied by choosing an Ansatz for (the local part of) $S_\text{cl}$ that is covariant on the radial cut-off slice.}
\begin{align}\label{eq:hj-eq-unexpanded}
\{S_\text{cl},S_\text{cl}\}-\mathcal{L}\ =\ 0
\end{align}
where the brackets are given by the following expression. Let $F$ and $G$ be two arbitrary phase-space functionals, then the brackets are defined as
\begin{align}\label{eq:hj-brackets}
\{F,G\}\ &\equiv\ \frac{-1}{(\sqrt{-\gamma})^2}\bigg[\left(\gamma_{ac}\gamma_{bd}-\frac1{d-1}\gamma_{ab}\gamma_{cd}\right)\frac{\delta F}{\delta \gamma_{ab}}\frac{\delta G}{\delta \gamma_{cd}}
+\frac12\gamma_{ab}\frac{\delta F}{\delta A_a}\frac{\delta G}{\delta A_b} \\
&\hspace{8cm}+\frac1{2m^2}D_a\frac{\delta F}{\delta A_a}\,D_b\frac{\delta G}{\delta A_b}\,\bigg]\nonumber,
\end{align}
where $\gamma_{ab}$ and $A_a$ are the induced fields pulled back onto the radial cut-off slice. The brackets \eqref{eq:hj-brackets} were introduced in \cite{deBoer:1999xf}; see also \cite{Baggio:2011cp}. It should be noted that these brackets are only introduced as a short-hand notation for the `kinetic' part of the Hamiltonian contraint; they are \emph{not} Poisson brackets (or any other type of special brackets).

We split up the on-shell action $S_\text{cl}$ into a local power-law divergent piece $S_\text{loc}$ and a non-local piece $W$ that is at most logarithmically divergent; $W$ is finite in the absence of anomalies. Using the split $S_\text{cl}=S_\text{loc}+W$, we may write the HJ equation as
\begin{align}\label{eq:hj-equation}
0\ =\ \{S_\text{loc},S_\text{loc}\}-\mathcal{L} + 2 \{S_\text{loc},W \} + \{W,W\}.
\end{align}
We define $\mathcal{H}_\text{loc}\equiv\{S_\text{loc},S_\text{loc}\}-\mathcal{L}$, and we require that the divergent terms cancel.\footnote{Here we refer to a term $F$ as ``divergent'' if $\lim_{r\to\infty}\sqrt{\gamma} F = \infty$.} It might happen that we get a finite remainder $\mathcal{H}_{\text{rem}}$ that cannot be removed by local counterterms. As a consequence, the HJ equations in the large $r$ limit, where we can ignore $\{W,W\}$, give:
\be
2\{S_{\text{loc}},W\} \approx -\mathcal{H}_{\text{rem}}.
\ee
The symbol ``$\approx$'' means an equality in the large $r$ limit. We will presently see that this is related to the anisotropic scaling anomaly in the dual non-relativistic field theory. First, we need to specify the boundary conditions in order to be able to identify which terms are divergent.

\subsection{Boundary conditions and anomaly}
\label{sec:boundaryconditions}
From the field theory side, we know that the volume form has a definite scaling weight $[\text{Vol}_d]=[dt\,d^{d-1}x]=z+d-1$. In the dual gravitational picture this weight is translated to a radial scaling, such that
\begin{align}\label{eq:vol-scaling-condition}
z+d-1\ =\ [\text{Vol}_d]\ =\ \frac{\partial_r\sqrt\gamma}{\sqrt\gamma}\ =\ \partial_r\log N+\frac12\partial_r\log\det h
\end{align}
Here, $N$ is the lapse function and $h_{ij}$ is the induced metric on a spatial slice of $\Sigma_r$. We assume that the spatial metric is of the form $h_{ij}=e^{2r}\hat h_{ij}$, where $\hat h_{ij}$ has a finite limit as $r \rightarrow\infty$. It then follows that the lapse scales as $N\sim e^{zr}$. This puts a restriction on the degrees of freedom contained in the metric. It implies in particular that we must turn off the off-diagonal mode in $\gamma_{ti}$ that scales as $e^{2zr}$; in terms of the linearized modes discussed in \cite{Ross:2009ar,Ross:2011gu,Baggio:2011cp}, one needs to kill the $c_{1i}$ mode. This naturally leads us to consider only deformations with a preferred time foliation, as suggested also in \cite{Ross:2011gu}.

Let us redefine $N$ and $h_{ij}$ to be the renormalized lapse and induced metric,\footnote{Again, we do not consider a shift $N^i$, as it can locally be removed by a foliation-preserving diffeomorphism.}
\begin{align}
\label{eqn:lifshitzconformalinfinity}
N & = \lim_{r \rightarrow\infty}e^{-zr}(-\gamma^{tt})^{-1/2},\\
h_{ij} & = \lim_{r \rightarrow\infty}e^{-2r}\gamma_{ij},
\end{align}
such that the renormalized volume form is given by $N\sqrt h=\lim_{r \rightarrow\infty} e^{-(z+d-1)r}\sqrt\gamma$. With these conditions, it is straightforward to see that:
\be
\label{eqn:anomaloustransformation}
2\sqrt{\gamma}\,\{S_{\text{loc}},W\} \approx 2z\, N^{2} \frac{\delta W}{\delta N^{2}} + 2 h_{ij} \frac{\delta W}{\delta h_{ij}} + z \hat{A}_{t} \frac{\delta W}{\delta \hat{A}_{t}} \approx -\sqrt{\gamma}\, \mathcal{H}_{\text{rem}},
\ee
where $\hat{A}_t = \lim_{r\rightarrow\infty}e^{-z r} A_t$. As noted in \cite{Ross:2011gu,Baggio:2011cp}, for $z\geq 2$ and $d=3$ the vector field becomes an irrelevant operator, which prevents us from defining the on-shell action non-perturbatively in the sources. In particular, the linearized analysis for $z=2$ shows the presence of a mode that shifts the value of $\alpha = A^{a}A_{a}$ and introduces logarithmic divergences in the metric:
\begin{align}
\gamma_{tt} &= -e^{4 r} \left(1+ \left(4c_3\,r+c_4\right) + \frac1{12}\left(4c_1-5c_2+4c_2\,r\right)\,e^{-4r}\right), \\
\gamma_{ij} &=  e^{2 r}\delta_{ij}\left(1 + \left(-2c_3\,r+c_5\right) + \frac1{24}\left(4c_1+5c_2+4c_2\,r\right)\,e^{-4r}\right),\nonumber\\
\alpha &= -1 + 2 c_{3}+ 2\left(c_1+c_2\,r\right)\,e^{-4r}.\nonumber
\end{align}
It is convenient to expand the counterterms in powers of $\alpha+1$, i.e.\ in fluctuations of $\alpha$ around its background value $\alpha_0=-1$. However, turning on the mode $c_{3}$ means that $(\alpha-\alpha_{0})^{n}$ is not suppressed as $r\to\infty$, and an infinite set of counterterms would be required.  Furthermore, the mode $c_{3}$ introduces logarithmic divergences in the metric sector, spoiling our definition of anisotropic conformal infinity \eqref{eqn:lifshitzconformalinfinity}. For this reason, we will turn off this mode\footnote{As shown in \cite{Baggio:2011cp}, it is possible to consistently impose this condition when higher order non-linear corrections are considered, and we believe there is no obstruction at the full non-linear level.} by setting $c_{3} = 0$. In particular notice the important relation:
\be
A_{t} = e^{2 r} N + \text{subleading},
\ee
where the subleading terms scale as $e^{-4r}$. In the frame field language of \cite{Ross:2011gu}, this corresponds to $\delta(A_{A}) = \delta(A_{a} e^{a}_{A}) = 0$. Notice that with these boundary conditions, $W$ becomes a functional of $N$ and $h$ only. Therefore we have:
\be
\frac{\partial W}{\partial N}\bigg|_{h = \text{const}} = \frac{\delta W}{\delta N} + \frac{\delta W}{\delta \hat{A}_{t}}\frac{\partial \hat{A}_{t}}{\partial N},
\ee
where the variations on the right are unconstrained, while the variation on the left represents the total variation of $W$ with respect to $N$. Therefore \eqref{eqn:anomaloustransformation} becomes:
\be
2 N \frac{\partial W }{\partial N} + 2 h_{ij} \frac{\delta W }{\delta h_{ij}} = -\sqrt{-\gamma}\,\mathcal{H}_\text{rem}.
\ee
By comparing with \eqref{eqn:traceanomaly}, we find that the anomaly is given by
\be\label{eq:def-holographic-anomaly}
\mathcal{A} =  -\lim_{r\to\infty} e^{4r}\, \mathcal{H}_{\text{rem}}.
\ee
This is the holographic anomaly in the HJ formalism. In the following we will use the aforementioned boundary conditions in order to identify possible divergent terms.

\subsection{Two-derivative counterterms and anomaly}
The Hamiltonian constraint at the level of two spacetime derivatives is given by
\begin{align}\label{eq:level-2-ham-constraint}
\mathcal{H}^{(2)}_\text{loc}\ &=\ \{S_\text{loc},S_\text{loc}\}^{(2)} -R+\frac14F_{ab}F^{ab}
\end{align}
Now, let us take the most general Ansatz for the on-shell action in such a way that $\mathcal{H}_a=0$ (i.e.\ covariant on $\Sigma_r$). At the non-derivative level, we thus take
\begin{align}\label{eq:ansatz-level0}
S_\text{loc}^{(0)}\ &=\ \int d^dx\sqrt{\gamma}\,\mathcal{F}_0(\alpha),
\end{align}
with $\mathcal{F}_0(\alpha)$ an arbitrary function of the Lorentz scalar $\alpha\equiv A_aA^a$. This case was analyzed in detail in \cite{Baggio:2011cp}, and does not lead to anomalies in the present case. At the level of two spacetime derivatives, we take\footnote{In order to get a general understanding of these terms, notice that the terms proportional to $\mathcal{F}_i$ with $i=1,2,3$ are of the schematic form $\gamma\gamma\, DA\,DA$, for $i=4,5,6,7$ they look like $AA\gamma\,DA\,DA$, and for $i=7$ is $AAAA\,DA\,DA$.}
\begin{align}\label{eq:ansatz-level2}
S_\text{loc}^{(2)}\ &=\ \int d^dx\sqrt{\gamma}\,\Big\{
\mathcal{F}_1\,R
+\mathcal{F}_2\,D_{a}A_{b}\,D^{a}A^{b}
+\mathcal{F}_3\,D_{a}A_{b}\,D^{b}A^{a}
+\mathcal{F}_4\,(D_aA^a)^2
+\mathcal{F}_5\,\partial^a\alpha\,\partial_a\alpha \nonumber\\&\quad
+\mathcal{F}_6\,A^bD_bA_a\,A^cD_cA^a
+\mathcal{F}_7\,A^a\partial_a\alpha\,D_bA^b
+\mathcal{F}_8\,A^aD_aA^b\,\partial_b\alpha 
+\mathcal{F}_9\,(A^a\partial_a\alpha)^2
\Big\},
\end{align}
where each coefficient $\mathcal{F}_i$ is a function of $\alpha$. We do not include terms that involve second derivatives, e.g.\ $\sim D_aD_bA_c$, since those can be expressed in terms of first-derivative terms by means of partial integration. Equation \eqref{eq:level-2-ham-constraint} can be solved straightforwardly by expanding the coefficient-functions around the Lifshitz background,
\begin{align}
\mathcal{F}_i(\alpha)\ &=\ \sum_{n\geq0}\,f_{i(n)}(\alpha-\alpha_0)^n.
\end{align}
The coefficients $f_{i(n)}$ are then found by solving the equation $\mathcal{H}^{(2)}_\text{loc}=0$ recursively order by order.

For the case of $z=2$ in $d=3$ boundary dimensions one finds a break-down of the recursive `descent' equations, i.e.\ $\mathcal{H}^{(2)}_\text{loc} \neq 0$ for any choice of coefficient functions $\mathcal{F}_{i}$. Such a break-down signals the presence of an anomalous remainder $\mathcal{H}_\text{rem}$ giving rise to the holographic anomaly. As explained in the introduction to this section, we should still be able to determine the coefficient of the \emph{divergent} part of the on-shell action in order to have a well-defined theory on the boundary. Doing so leads to the following remainder:\footnote{
To see how this comes about, one can can look at the solution $S_\text{loc}=\int d^dx\sqrt{-\gamma}\,\mathcal{L}_\text{loc}$ for values of the dynamical exponent close to the critical value, $z\approx2$. One finds a simple pole at $z=2$,
$$
\mathcal{L}^{(2)}_\text{loc}\ =\ \frac1{z-2}\Big(-D_{a}A_{b}\,D^bA^a + \frac{1}{2} (D_{a} A^{a})^{2}\Big) + O\left((z-2)^0\right),
$$
where the residue is simply the two-derivative remainder \eqref{eq:H2rem}.
}
\begin{align}\label{eq:H2rem}
\mathcal{H}^{(2)}_\text{rem}\ =\ -D_{a}A_{b}\,D^bA^a + \frac{1}{2} (D_{a} A^{a})^{2},
\end{align}
which is obviously affected by the ambiguity of adding finite local counterterms to the action. Using the boundary conditions described in \ref{sec:boundaryconditions} and the definition of the holographic anomaly \eqref{eq:def-holographic-anomaly}, we find the following two-derivative contribution to the anomaly:
\be
\mathcal{A}\ =\ \frac{\ell}{64\pi\,G} \,\frac{1}{N^{2}} \left(  h^{ij}h^{kl}\,\dot{h}_{ik}  \dot{h}_{jl} -\frac12(h^{ij}\dot{h}_{ij})^{2}\right)+\ldots,
\ee
where the ellipses denote possible four-derivative contributions. The above anomaly is indeed of the form \eqref{eqn:genanomaly} with:
\be
C_{1} = \frac1{128\pi}\,\frac{2\ell}G
\ee
We reinstated the four-dimensional Newton's constant $G$ and the curvature length scale $\ell$. Also, $\dot{h}_{ij}$ is generically related to the extrinsic curvature by $K_{ij}=(\dot{h}_{ij}-\nabla_iN_j-\nabla_jN_i)/2N$ (and we had set the shift $N_i$ to zero).\medskip

\subsection{Four-derivative anomaly}
One may repeat the above steps at the level of four derivatives.\footnote{One does not expect to find anomalous contributions that contain three derivatives (one time and two spatial), since terms that involve an odd number of time-derivatives are not invariant under time-reversal.} The four-derivative Ansatz is
\begin{align}
S^{(4)}_\text{loc}\ =\ \int d^dx\sqrt\gamma\,\Big( \mathcal{G}_1\,R^2+\mathcal{G}_2\,R_{ab}R^{ab}+\mathcal{G}_3\,R_{abcd}R^{abcd}+\mathcal{G}_4\,\square R \Big)+\ldots
\end{align}
The ellipses denote terms that involve the Proca field $A_a$. All the terms that appear at this level are finite for our choice of boundary conditions, which means that they can only contribute with trivial total derivatives to the anomaly. In this case, we find the remainder
\begin{align}
\mathcal{H}^{(4)}_\text{rem}\ =\ \frac18 R^2-\frac14R_{ab}R^{ab} + \ldots
\end{align}
where the ellipses denote once again terms that involve $A_{a}$.
Writing this in terms of the two-dimensional Ricci tensor gives:
\begin{align}
\label{eq:lif-remainder-4deriv}
\mathcal{H}^{(4)}_\text{rem}\ =\ \frac14\,{}^{2\!\!}R_{ij}{}^{2\!\!}R^{ij}-\frac18\, ({}^{2\!\!}R)^2+\ldots,
\end{align}
where these ${}^{2\!\!}R$ and ${}^{2\!\!}R_{ij}$ are the \emph{two}-dimensional Ricci scalar and tensor and we have not written down terms that involve derivatives acting on $N$. We can use the off-shell identity that relates the Ricci tensor to the Ricci scalar, ${}^{2\!\!}R_{ij}=\frac12{}^{2\!\!}R\,h_{ij}$, which is specific to two dimensions. When we plug this into \eqref{eq:lif-remainder-4deriv}, we find that the Ricci-squared terms cancel and we have (equation \eqref{eqn:genanomaly}):
\be
C_{2} = 0,
\ee
which agrees with the field theory computation.

Notice that while we were able to extract the coefficient $C_{2}$, we have not performed a complete analysis of the counterterms at the four-derivative level, which would be rather involved. Nevertheless, the complete answer has been computed using the results of \cite{Ross:2011gu} in \cite{Griffin:2011xs}, in perfect agreement with our result $C_{2}=0$. In conclusion, the holographic anomaly is given by
\be
\mathcal{A}\ =\ \frac{\ell}{64\pi\,G} \,\frac{1}{N^{2}} \left(  h^{ij}h^{kl}\,\dot{h}_{ik}  \dot{h}_{jl} -\frac12(h^{ij}\dot{h}_{ij})^{2}\right).
\ee
%



\section{Discussion and conclusions}

In this paper we computed the anisotropic scaling anomaly of two Lifshitz theories, 
one defined using a standard field theory quantization of an explicit classical action \eqref{eqn:lifshitzmodel}, 
the other defined using the holographic correspondence. 
A precise definition of Lifshitz holography is still lacking, and a microscopic definition of the strongly
coupled field theory is not known. It is therefore a priori not very meaningful to compare the two anomalies.
Nevertheless, we found that the anomalies are quite similar. In both cases there are two possible central
charges of which one vanishes, and as a consequence the two anomalies are directly proportional to each other.
The ratio of the two anomalies, in the conventions used in this paper, is $2\ell/G$, with $\ell$ the
curvature radius of the Lifshitz spacetime and $G$ the 4d Newton constant. It would be interesting
to evaluate this quantity in explicit string theory embeddings of Lifshitz spacetimes to see how
it scales with the various integer fluxes, as this will provide some measure of the effective number
of degrees of freedom of the dual field theory. 

It is quite mysterious that the conformal anomaly only involves time derivatives, it is even mysterious
that there exists a conformal anomaly at all. According to \cite{Adams:2008zk}, the dynamical critical
exponent is in general renormalized, and as soon as $z=2+\epsilon$ a conformal anomaly can no longer
be written down. So either there is some unknown mechanism that protects the value of $z=2$, or the conformal
anomaly can be removed in the full quantum theory. In the latter case, one would be in the peculiar
situation that one would need to include counterterms that diverge in the classical limit. Further work
will be required to clarify this issue.

It is  also clearly of interest to explore other systems with anisotropic scale invariance to examine
whether the conformal anomaly is still of the same form. In particular, whenever one has a Lifshitz solution
in a theory with Chern-Simons type terms, time reversal symmetry is broken and it is logically possible
to have contributions with an odd number of time derivatives to the conformal anomaly. It is in principle
straightforward to extend the analysis in appendix~A to determine whether there are non-trivial terms
of this type and we leave this as an exercise. 

As mentioned before, one of the main uses of the conformal anomaly is that it is a relatively simple 
property of a field theory which sometimes gives rise to certain universal properties. For example,
in the relativistic case, in $d=2$ the conformal anomaly completely fixes the free energy
at high temperatures, and the central charges also control the logarithmic divergences 
in the entanglement entropy in $d=2,4$. Whether similar universal properties also exist for
non-relativistic field theories is an interesting open problem that we hope to come back to in
the future.

\section*{Acknowledgements}
We would like thank Juan Jottar, Geoffrey Comp\`ere, Balt van Rees and Simon Ross for fruitful discussions. We would also like to thank the participants of the 
Meeting on Holography at Finite Density held from 
November 16-18, 2011 at the APC in Paris for useful discussions. 
This work is part of the research programme of the Foundation for Fundamental Research on Matter (FOM), 
which is part of the Netherlands Organisation for Scientific Research (NWO).


\appendix


\section{Classification of possible terms in the anomaly}
\label{sec:partialintegration}
In this appendix we explore to what extent it is possible to remove total derivatives from the anomaly. This is achieved by adding appropriate scale invariant counterterms to the action that are not invariant under \emph{local} scale transformations. Clearly, we can discuss the two-derivative and the four-derivative terms separately. Let us start with the former; there are only three possible scale-invariant terms that we can construct with two time derivatives:
\be
h^{ij} \frac{1}{N} \partial_{t} (\frac{1}{N}  \partial_{t} h_{ij}), \qquad \frac{1}{N^{2}}(h^{ij} \dot{h}_{ij})^{2}, \qquad h^{ij}\dot{h}_{jk} h^{kl}\dot{h}_{li}.
\ee
It is straightforward to see that the two combinations
\be
h^{ij} \frac{1}{N} \partial_{t} (\frac{1}{N}  \partial_{t} h_{ij}), \qquad \frac{1}{N^{2}}\left(h^{ij} h^{kl}\,\dot{h}_{ik}\dot{h}_{jl} - \frac12 (h^{ij} \dot{h}_{ij})^{2} \right),
\ee
are invariant under \emph{local} scale transformations (up to total derivatives). These two terms are related by partial integration, and we now show that it is indeed possible to ``partially integrate'' inside the anomaly by adding an appropriate counterterm to the action. The most general form of the anomaly at the two derivative level is:
\begin{equation}
\delta W = \int \delta\rho \left\{a \frac{1}{N} h^{ij}\partial_{t} (\frac{1}{N}  \dot{h}_{ij}) + b \frac{1}{N^{2}}\left(h^{ij} h^{kl}\,\dot{h}_{ik}\dot{h}_{jl} - \frac12 (h^{ij} \dot{h}_{ij})^{2} \right)\right\}.
\end{equation}
The presence of the factor $\delta\rho$ prevents us from doing partial integration directly. Let us add the following counterterm to the action:
\be
W' = W + c \int N\sqrt{h} \frac{1}{N^{2}} (h^{ij} \dot{h}_{ij})^{2}.
\ee
It is then easy to check that
\be
\delta W' =  \int \delta\rho \left\{(a-8c) \frac{1}{N} h^{ij}\partial_{t} \frac{1}{N} \dot{h}_{ij} + (b+8c)\frac{1}{N^{2}}\left(h^{ij} h^{kl}\,\dot{h}_{ik}\dot{h}_{jl} - \frac12 (h^{ij} \dot{h}_{ij})^{2} \right)\right\}.
\ee
Therefore we can pick $c=a/8$ and get rid of the first term, which is tantamount to integrating by parts, or discarding total derivatives in the anomaly. For instance, in the field theory analysis, we went from \eqref{aux1} to \eqref{aux1-after-part-intgr} using this procedure. In particular, we had $a=1/48\pi$ and $b=-5/384\pi$, such that $b+8c=a+b=1/128\pi$.\footnote{Bear in mind that there was an extra factor $-2$ coming from the relation between $\tilde{a}_2(\delta\rho,D)$ and the integrated anomaly, cf.\ \eqref{eqn:anomtrans}.}

Let us now consider the four derivative level. In this case we are interested in terms of the form $\nabla_{i} J^{i}$ in the anomaly. We ask ourselves to what extent it is possible to remove them by adding local counterterms $G$ to the action. Both the total derivatives and the local counterterms must be scale invariant, therefore there is only a finite number of them. Let us choose a basis:
\begin{align}
J^{i}_{a} & \qquad a = 1,\ldots,N\\
G_{b} & \qquad b = 1,\ldots,M.
\end{align}
The Weyl variation of a linear combination $\sum_{b} q_{b}\, G_{b}$ can be written, after partial integration, as:
\begin{align}\label{eq:def-weyl-operator}
\delta \sum_{b}q_{b}\, G_{b} = \frac\omega{N} \sum_{ab} M_{ab}\,q_{b}\, \nabla_{i}J^{i}_{a},
\end{align}
If the variation of the effective action reads:
\be
\delta W = \int\sqrt h\, \omega \left(\mathcal{A} + \sum_{a} c_{a} \nabla_{i}J^{i}_{a}\right),
\ee
we can get rid of the total derivatives if we can solve the system of linear equations:
\be
M_{ab}\, q_{b} = c_{a}.
\ee
If we are to remove all the possible total derivatives that can appear, the number of rows $N$ of the matrix $M_{ab}$ must be less than or equal to the number of columns $M$, and the rank of the matrix should be maximal. It is easy to check that there are 6 possible functionally independent scale invariant currents $J^{i}$, and we choose the following basis:
\be
\begin{array}{lll}
J^{i}_{1} = N \partial^{i} R  \quad & J^{i}_{2} = (\partial^{i} N) R \quad & J^{i}_{3} = (\partial^{i} N) (\frac{1}{N} \partial_{j} N) (\frac{1}{N} \partial^{j} N)\\
J^{i}_{4} = (\partial^{i} N) (\frac{1}{N} \Delta N) \quad & J^{i}_{5} = (\partial^{j} N) (\frac{1}{N} \nabla_{j} \partial^{i} N) \quad & J^{i}_{6} = \partial^{i} \Delta N.
\end{array}
\ee
Analogously, there are 12 functionally independent scale invariant counterterms, and we choose the basis:
\be
\begin{array}{lll}
G_{1} = R^{2} \quad & G_{2} =  \Delta R  \quad & G_{3} = (\frac{1}{N} \Delta N) R \\[10pt]
G_{4} = (\frac{1}{N} \partial_{i} N) (\frac{1}{N} \partial^{i}N) R \quad & G_{5} =  ((\frac{1}{N} \partial_{i} N) (\frac{1}{N} \partial^{i}N))^{2} \quad & G_{6} =  (\frac{1}{N} \partial_{i} N) (\frac{1}{N} \partial^{i}N) (\frac{1}{N} \Delta N) \\[10pt]
G_{7} = (\frac{1}{N} \Delta N)^{2} \quad & G_{8} = (\frac{1}{N} \partial^{i} N)(\frac{1}{N} \partial^{j} N) (\frac{1}{N} \nabla_{i}\partial_{j} N)\quad & G_{9} = (\frac{1}{N}\partial^{i}N) \frac{1}{N}\partial_{i} \Delta N\\[10pt]
G_{10} = \frac{1}{N} \nabla_{i}\partial_{j} N \frac{1}{N} \nabla^{i} \partial^{j} N & G_{11} = \frac{1}{N}\Delta^{2} N & G_{12} = \frac{1}{N} \partial^{i}N \partial_{i} R
\end{array}
\ee
While we have many more possible counterterms than currents, it is important to stress that not all the counterterms are independent, since we can always partially integrate inside the action. This means that some linear combinations of counterterms will have the same Weyl transformation. Furthermore, there can be Weyl invariant combinations of counterterms that do not help in removing total derivatives from the anomaly.

By taking the Weyl variation of the 12 terms $G_{b}$, it is straightforward to compute the matrix $M$, which is given by:
\be
M_{ab} = \left(
\begin{array}{cccccccccccc}
-4 & 2 & 2 & 0 & 0 & 0 & 0 & 0 & 0 & 0 & 0 & -2\\
-4 & -2 & -2 & -4 & 0 & 0 & 0 & 0 & 0 & 2 & 0 & 2\\
0 & 0 & 0 & 2 & -8 & -6 & 0 & -5 & -6 & 0 & 0 & 0\\
0 & 0 & 0 & 0 & 0 & -4 & -8 & 2 & 4 & -2 & 0 & 0\\
0 & 0 & 0 & -4 & 0 & 4 & 0 & -2 & 4 & -4 & 0 & 0\\
0 & -2 & -2 & 0 & 0 & 0 & 4 & 0 & -4 & 4 & 0 & 2
\end{array}
\right)
\ee
It is easily checked that $M$ does \emph{not} have maximal rank (which would be 6), but it has rank 5. In fact, $M$ has a 7 dimensional space of null vectors, which is spanned by the 6 total derivatives $\nabla_{i}J^{i}$ and a Weyl invariant term:
\be
\delta\int \sqrt h\,\nabla_iJ^i_a = 0, \qquad \delta\int N\sqrt h\,\Big( R+\frac1N\Delta N-\frac1{N^2}\partial_iN \partial^iN \Big)^2\ =\ 0.
\ee
Since the rank of $M$ is 5, the Weyl variation of the most general counterterm spans a 5 dimensional subspace of the 6 dimensional space generated by $c_a\,\nabla_i J^i_a$. That means that we can find an orthonormal basis (with respect to the usual Euclidean scalar product $\delta_{ab}$) for the currents where 5 are trivial (i.e. removable by counterterms) and 1 is non-trivial. In other words, we look for 5 vectors $e_{a}$ such that $e_a=M_{ab}q_b$ admits a solution. If we now take $u_a$ to be the null vector of the transpose of $M_{ab}$, it is obviously orthogonal to all the $e_{a}$ since $u_{a} e_{a} = e_{a}M_{ab}q_{b} = 0$. We define the non-trivial current $\mathcal{J}^{i}$ to be:
\be
\label{eqn:nontrivialcurrent}
\mathcal{J}^{i} = u_{a} J^{i}_{a} = J^i_1-J^i_2+J^i_4+J^i_5+2J^i_6.
\ee
However, we will presently show that this current does \emph{not} obey the Wess--Zumino consistency condition, therefore it cannot appear in the anomaly.

\subsection*{Wess--Zumino consistency condition and $\mathcal{J}^i$}
The goal of this section is to figure out whether all possible terms that we found above satisfy the Wess--Zumino consistency conditions. To this end, we shall compute the quantities
\begin{align}
\Omega_a\ &\equiv\ \delta_1\int d^2x\sqrt{h}\,\omega_2\nabla_iJ^i_a -\delta_2\int d^2x\sqrt{h}\,\omega_1\nabla_iJ^i_a \\
&=\ \int d^2x\,\delta_2\big(\sqrt{h}\,J^i_a\big)\,\partial_i\omega_1-\int d^2x\,\delta_1\big(\sqrt{h}\,J^i_a\big)\,\partial_i\omega_2\label{eq:Omega}
\end{align}
for each $a=1,..,6$. The main idea of this analysis is to find all possible linear combinations of the $\Omega$'s such that
\begin{align}
\sum_{a=1}^6c_a\Omega_a\ =\ 0
\end{align}
If the vector space spanned by the vectors $\{\vec{c}\}$ is six dimensional, all $J_a^i$'s are Wess--Zumino-consistent. If, on the other hand, this vector space is five-dimensional then we must conclude that one of the $J_a^i$'s is inconsistent. Since we already know that five currents can be generated by varying appropriate local scale invariant terms, these are manifestly consistent. Therefore the inconsistent current, if present, must be the non-trivial current of equation \eqref{eqn:nontrivialcurrent}.

The way we shall carry out this computation is by first computing the first term in \eqref{eq:Omega}. The second term in \eqref{eq:Omega} is then obtained from the first one by replacing the derivatives that act on $\omega_1$ for derivatives that act on $\omega_2$ by means of partial integration.

We shall start with $\Omega_1$. The first term in  \eqref{eq:Omega} is\footnote{For notational clarity, notice that the variation differs by a factor of two compared to before. For instance, $h_{ij}\to e^{\omega}h_{ij}$ rather than  $h_{ij}\to e^{2\omega}h_{ij}$.}
\begin{align}
\delta_2 \big(\sqrt{h}\,J_1^i\big)\,\partial_i\omega_1\ &=\ \sqrt{h}\left( -\partial^i\omega_2\,NR-\partial^i\Delta\omega_2\,N \right)\,\partial_i\omega_1
\end{align}
The second term is then
\begin{align}
\delta_1 \big(\sqrt{h}\,J_1^i\big)\,\partial_i\omega_2\ &=\ \sqrt{h}\left( -\partial^i\omega_1\,NR-\partial^i\Delta\omega_1\,N \right)\,\partial_i\omega_2 \\
&=\ \sqrt{h}\left( -\partial^i\omega_2\,NR-\nabla^i \nabla^j\big(\partial_j\omega_2\,N\big) \right)\,\partial_i\omega_1 \\
&=\ \sqrt{h}\big( -\partial^i\omega_2\,NR - \partial^i\Delta\omega_2\,N\\
&\hspace{2cm} - \Delta\omega_2\,\partial^iN - \partial^i\big( \partial_j\omega_2\,\partial^jN\big) \big)\,\partial_i\omega_1
\end{align}
so that
\begin{align}
\Omega_1\ =\ \int d^2x\sqrt{h}\,\left( \Delta\omega_2\,\partial^iN + \partial^i\big( \partial_j\omega_2\,\partial^jN\big) \right) \,\partial_i\omega_1
\end{align}
Similarly, from $J_2^i$:
\begin{align}
\delta_2 \big(\sqrt{h}\,J_2^i\big)\,\partial_i\omega_1\ &=\ \sqrt{h}\left( \partial^i\omega_2\,NR -\Delta\omega_2\,\partial^iN\right)\,\partial_i\omega_1\\
\delta_1 \big(\sqrt{h}\,J_2^i\big)\,\partial_i\omega_2\ &=\ \sqrt{h}\left( \partial^i\omega_1\,NR - \Delta\omega_1\,\partial^iN \right)\,\partial_i\omega_2 \nonumber\\
&=\ \sqrt{h}\left( \partial^i\omega_2\,NR + \partial^i\big(\partial_j\omega_2\,\partial^jN\big) \right)\,\partial_i\omega_1 \\
\Omega_2\ &=\ -\int d^2x\sqrt{h}\,\left( \Delta\omega_2\,\partial^iN + \partial^i\big( \partial_j\omega_2\,\partial^jN\big) \right) \,\partial_i\omega_1
\end{align}
From $J_3^i$:
\begin{align}
\delta_2 \big(\sqrt{h}\,J_3^i\big)\,\partial_i\omega_1\ &=\ \sqrt{h}\left( \partial^i\omega_2\,\partial_jN\,\partial^jN + 2\partial_j\omega_2\,\partial^iN\,\partial^jN\right)\,\partial_i\omega_1\\
\delta_1 \big(\sqrt{h}\,J_3^i\big)\,\partial_i\omega_2\ &=\ \sqrt{h}\left( \partial^i\omega_1\,\partial_jN\,\partial^jN + 2\partial_j\omega_1\,\partial^iN\,\partial^jN\right)\,\partial_i\omega_2 \nonumber\\
&=\ \sqrt{h}\left( \partial^i\omega_2\,\partial_jN\,\partial^jN + 2\partial_j\omega_2\,\partial^jN\,\partial^iN\right)\,\partial_i\omega_1 \\
\Omega_3\ &=\ 0
\end{align}
From $J_4^i$:
\begin{align}
\delta_2 \big(\sqrt{h}\,J_4^i\big)\,\partial_i\omega_1\ &=\ \sqrt{h}\left( \partial^i\omega_2\,\Delta N+2\partial_j\omega_2\,\tfrac1N \partial^iN \partial^jN+\Delta \omega_2\,\partial^iN \right)\,\partial_i\omega_1\\
\delta_1 \big(\sqrt{h}\,J_4^i\big)\,\partial_i\omega_2\ &=\ \sqrt{h}\left( \partial^i\omega_1\,\Delta N+2\partial_j\omega_1\,\tfrac1N \partial^iN \partial^jN+\Delta \omega_1\,\partial^iN \right)\,\partial_i\omega_2 \nonumber\\
&=\ \sqrt{h}\left( \partial^i\omega_2\,\Delta N+2\partial_j\omega_2\,\tfrac1N \partial^jN \partial^iN - \partial^i\big( \partial_j\omega_2\,\partial^jN \big) \right)\,\partial_i\omega_1 \\
\Omega_4\ &=\ \int d^2x\sqrt{h}\,\left( \Delta\omega_2\,\partial^iN + \partial^i\big( \partial_j\omega_2\,\partial^jN\big) \right) \,\partial_i\omega_1
\end{align}
From $J_5^i$:
\begin{align}
\delta_2 \big(\sqrt{h}\,J_5^i\big)\,\partial_i\omega_1\ &=\ \sqrt{h}\left( \partial_j\omega_2\,\nabla^i \partial^jN+\partial^i\omega_2\,\tfrac1N \partial^jN \partial_jN + \nabla_j \partial^i\omega_2\,\partial^jN \right)\,\partial_i\omega_1\\
\delta_1 \big(\sqrt{h}\,J_5^i\big)\,\partial_i\omega_2\ &=\ \sqrt{h}\left( \partial_j\omega_1\,\nabla^i \partial^jN+\partial^i\omega_1\,\tfrac1N \partial^jN \partial_jN + \nabla_j \partial^i\omega_1\,\partial^jN \right)\,\partial_i\omega_2 \nonumber\\
&=\ \sqrt{h}\left( \partial_j\omega_2\,\nabla^i \partial^jN+\partial^i\omega_2\,\tfrac1N \partial^jN \partial_jN -\nabla_j \big( \partial^{(i}\omega_1\,\partial^{j)}N \big) \right)\,\partial_i\omega_1 \\
\Omega_5\ &=\ \int d^2x\sqrt{h}\,\left( \Delta\omega_2\,\partial^iN + \partial^i\big(\partial_j\omega_2\, \partial^jN\big) \right) \,\partial_i\omega_1
\end{align}
From $J_6^i$:
\begin{align}
\delta_2 \big(\sqrt{h}\,J_6^i\big)\,\partial_i\omega_1\ &=\ \sqrt{h}\,\partial^i\!\left( 2\partial_j\omega_2\,\partial^jN +\Delta\omega_2\,N \right)\,\partial_i\omega_1 \nonumber\\
&=\ \sqrt{h}\,\left( \partial^i\Delta\omega_2\,N + \Delta\omega_2\,\partial^iN + 2\partial^i\big(\partial_j\omega_2\, \partial^jN\big) \right) \,\partial_i\omega_1\\
\delta_1 \big(\sqrt{h}\,J_6^i\big)\,\partial_i\omega_2\ &=\ \sqrt{h}\,\partial^i\!\left( 2\partial_j\omega_1\,\partial^jN +\Delta\omega_1\,N \right)\,\partial_i\omega_2 \nonumber\\
&=\ \sqrt{h}\left( \partial^i\Delta\omega_2\,N -\Delta\omega_2\,\partial^iN \right)\,\partial_i\omega_1 \\
\Omega_6\ &=\ 2\int d^2x\sqrt{h}\,\left( \Delta\omega_2\,\partial^iN + \partial^i\big(\partial_j\omega_2\, \partial^jN\big) \right) \,\partial_i\omega_1 
\end{align}

We thus find that each $\Omega_a$ is a multiple of
\begin{align}
\int d^2x\sqrt{h}\,\left( \Delta\omega_2\,\partial^iN + \partial^i\big(\partial_j\omega_2\, \partial^jN\big) \right) \,\partial_i\omega_1,
\end{align}
which means that there is one linear combination that does \emph{not} satisfy the Wess--Zumino consistency conditions. In other words, all but one of the six $J^i_a$'s can be made consistent. Since we have already found that five of the six $J^i_a$'s can be canceled by variations of local terms, the one that cannot be canceled (which we called $\mathcal{J}^i$) must be inconsistent. We can make this more precise by noticing that the consistency equation
\begin{align}
c_1-c_2+c_4+c_5+2c_6\ =\ 0
\end{align}
describes a five-dimensional hypersurface of consistent linear combinations $c_aJ^i_a$. The set of all such $c_a$-vectors can be defined as those that are orthogonal to the \emph{inconsistent} vector, $v_a$ say, such that $c_av_a=0$. The inconsistent vector is
\begin{align}
\vec{v}\ =\ \begin{pmatrix} 1& -1& 0& 1& 1& 2 \end{pmatrix}
\end{align}
As a consistency check on our computations, notice that this is precisely the five-dimensional hypersurface that we mentioned above, which may be defined as all vectors that are orthogonal to $u_a$ (as defined in \eqref{eqn:nontrivialcurrent}). Namely, the vector $u_a$ is the same as the inconsistent vector, i.e.\ $u_a=v_a$. The fact that $\mathcal{J}^i$ does not satisfy the Wess--Zumino condition means that it cannot appear as the variation of either local or non-local terms. The fact that there are precisely five total-derivative terms in the anomaly, all of which can be canceled by variations of local terms, was also noted in \cite{Griffin:2011xs}.

\textbf{Note:} In an earlier version of this paper, we had claimed that there would be three instead of two possible contributions to the Lifshitz scaling anomaly, one of which was said to be proportional to $\int\sqrt{h}\,\nabla_i \mathcal{J}^i$. This erroneous conclusion has been corrected in the present version of this paper.

\section{Role of the massive vector on the field theory side}
In this section we explore the conformal invariance of the field theory Lifshitz model from a different perspective. In particular, we will show that a preferred timelike vector $n^{\mu}$ plays a very similar role to the vector field $A^{\mu}$ appearing in the bulk.

Our set-up is the following three-dimensional scalar model with critical exponent $z=2$ \cite{Ardonne:2003wa},
\begin{align}
S\ =\ \int d^2x\,dt\,\mathcal{L}\ =\ \frac12\int d^2x\,dt \Big( \dot\phi^2-(\Delta\phi)^2 \Big).
\end{align}
The operator $\Delta$ is the spatial Laplacian $\Delta=\delta^{ij}\partial_i\partial_j$ and the dot denotes differentiation with respect to (imaginary) time, $\dot\phi=\partial_t\phi$. The Noether current density $(J_a)^b$ corresponding to the infinitesimal diffeomorphism $x^a\mapsto x^a+\varepsilon^a$ is given via the usual definition\footnote{We use the notation $x^t=t$, i.e.\ the index $a$ runs over $a=t,1,2$.}
\begin{align}
\delta_\varepsilon S\ =\ \int d^2x\,dt\,(J_a)^b\,\partial_b \varepsilon^a,
\end{align}
The current $(J_t)^a$ generates time reparametrizations and $(J_i)^a$ generates the spatial ones; their components are given by
\begin{align}
(J_t)^t\ &=\ -\tfrac12\dot\phi^2-\tfrac12(\Delta\phi)^2 \\
(J_t)^i\ &=\ \partial^i\dot\phi\,\Delta\phi-\dot\phi\,\partial^i\Delta\phi \\
(J_i)^t\ &=\ -\dot\phi\,\partial_i\phi \\
(J_i)^j\ &=\ \delta^j_i\,\mathcal{L}+\partial_i\partial^j \phi\,\Delta\phi-\partial_i\phi\,\partial^j\Delta\phi\label{eq:Tij-noether}
\end{align}
where $\partial^i=\delta^{ij}\partial_j$. One thing we see here is that $(J_t)^t=-\mathcal{E}$, where $\mathcal{E}$ is the Hamiltonian/energy density. The conservation law reads
\begin{align}\label{eq:conservation-law}
\partial_b(J_a)^b\ =\ -\partial_a\phi\,(\ddot\phi+\Delta^2\phi)\ \approx\ 0
\end{align}
The symbol $\approx$ denotes weak equality, i.e.\ equality up to terms that vanish on shell. The `gauge' parameter that generates the Lifshitz scaling is $\varepsilon^t=2\varepsilon\,t$ and $\varepsilon^i=\varepsilon\,x^i$ ($\varepsilon$ is just a small real number). The condition for scale invariance is
\begin{align}
2\,(J_t)^t+(J_i)^i \ =\ \partial_i\left( -2\partial^i\phi\,\Delta\phi \right)
\end{align}
whose right-hand side is not zero but a total divergence. The conserved current $\mathcal{S}^a$ associated to scale invariance of the theory is
\begin{align}
\mathcal{S}^t\ &=\ 2t(J_t)^t+x^i(J_i)^t &
\mathcal{S}^i\ &=\ 2t(J_t)^i+x^j(J_j)^i
\end{align}
Note that we cannot interpret the $J$'s as comprising an energy momentum tensor, since it would be far from being symmetric.

If we couple the Lifshitz model to $N$ and $h_{ij}$, we can easily
write down the condition for conformal invariance, however the relation
between the bulk (with its complete metric and the extra gauge field)
and the field theory model is rather obscure. Clearly, the bulk
metric does not couple to the energy momentum tensor of the
field theory as defined through $J_t$ and $J_i$, since that tensor
is not even symmetric.

So we will now make a more precise proposal about the relation between
the two. We introduce a three-dimensional metric $g_{\mu\nu}$ and
a unit timelike vector $n_a$ so that $n_a n^a=-1$. 
Define the projector $h_a{}^b=g_a{}^b+n_a n^b$, which is orthogonal to $n^a$, and
\be
\Delta\phi\equiv  \partial_a (h^{ab}\partial_b\phi)
+\frac{1}{2} h^{ab}h^{cd}\, \partial_{a}\phi \,\partial_{b} h_{cd}
\ee
then we can couple the Lifshitz model to $h_{ab}$ and $n_{a}$ via
the covariant action
\be \label{aux112}
S = \int d^2x\,dt\sqrt{-g} \left( (n^{a} \partial_{a}\phi)^2 - (\Delta \phi)^2\right).
\ee
This action is conformally invariant under 
\be \label{aux111}
\delta n_{a} = 2\omega\, n_{a},\qquad
\delta h_{ab} = 2\omega\, h_{ab}
\ee
This is why it is useful to introduce $h_a{}^b$, since the three-dimensional metric itself would transform as $\delta g_{ab} = -4 \omega\, n_{a} n_{b} + 2\omega\, h_{ab}$ (using the completeness relation $g_{ab}=-n_a n_b+h_{ab}$). Of course, all of this is not very profound. We have merely replaced the spatial metric $h_{ij}$ 
by the projection of the metric in the plane perpendicular to unit normal $n_{a}$. 

The claim is that (\ref{aux112}) describes the coupling of the Lifshitz model
to a metric and a gauge field in exactly the same way as one would expect
from the bulk description.  

With this fully covariant action, we can define a symmetric "stress tensor"
by varying it with respect to $g_{ab}$. Due to the presence of $n_{a}$,
this stess-tensor is not conserved though. The precise equation that
expresses general covariance of the theory reads
\be
2 D_{b} \frac{\delta S}{\delta g_{ab}} = 
(D^{a} n_{b}) \frac{\delta S}{\delta n_{b}} -
D_{b} \left( n^{a} \frac{\delta S}{\delta n_{b}}\right),
\ee 
where on the left hand side we recognize the covariant derivative
of the stress-tensor. The background field $n_{a}$ is the quantity
that breaks the general covariance of the theory, which explains
why this equation has a right-hand side.

In view of (\ref{aux111}), to write the conformal anomaly in covariant variables,
we also need a variation in terms of $n_{a}$
\be
\sqrt{-g}\,{\cal A} = ( -4n_{a} n_{b} + 2h_{ab})\frac{\delta S}{\delta g_{ab}}
 + 2n_{a} \frac{\delta S}{\delta n_{a}}.
\ee
This is exactly the same as the bulk equation with $n_{a}$ playing the role
of the asympotic value of $A_{a}$. When we choose $n_t=N$, $g_{tt}=-N^2$, $g_{ti}=0$ and $g_{ij}=h_{ij}$, the conformal anomaly becomes the expression we have been using all along.

\section{Alternative computation of $C_{2}$}
\label{sec:alternative}
In this section we provide an alternative computation of $C_{2}$. Since the structure multiplying $C_{2}$ contains $R$, we can take $N=1$ and assume that
$h_{ij}$ does not depend on $t$ but does depend on $x^i$. With these assumptions, the $\omega$ integral separates out, yielding a factor
of $\sqrt{\pi/\epsilon}$. What is left is to study the operator $\exp(-\epsilon \nabla^2)$.
Now we can roughly think of the standard heat kernel expansion as the Laplace
transform of the spectral density. So by taking the inverse Laplace
transform we can reconstruct the spectral density. The inverse
transform of $\epsilon^a$ is $s^{-1-a}/\Gamma(-a)$. Next, we can integrate
this against $\exp(-\epsilon s^2)$ to obtain 
\be
\frac{\epsilon^{a/2} \Gamma(-a/2)}{2\Gamma(-a)}.
\ee
This suggests that if the operator $\nabla$ has heat kernel expansion
\be
\sum_{n \geq -1} \epsilon^{n} L_n
\ee
then $\nabla^2$ has the expansion
\be
\sum_{n\geq -1} \epsilon^{n/2} \frac{\Gamma(-n/2)}{2\Gamma(-n)} L_n.
\ee
The term with $n=1$, which would contribute to the anomaly, vanishes due to the Gamma function. Therefore the coefficient of the $R^{2}$ term vanishes\footnote{See also \cite{gilkey1980spectral} for a rigorous proof of this statement.}. We conclude that the coefficient $C_{2}$ in \eqref{eqn:genanomaly} vanishes as well.\\
Notice that while this method is simpler than a direct computation, it is not powerful enough to determine the total derivative terms.

\section{Divergent terms in the heat-kernel expansion}
The methods of section \ref{sec:heatkernel} allow us to compute the divergent part of the heat-kernel expansion for the model considered in equation \eqref{eqn:lifshitzmodel}. We present the result here for completeness:
\be
K(\epsilon,f,D) \sim \frac{1}{\epsilon} \tilde a_{0}(f,D) + \frac{1}{\sqrt{\epsilon}} \tilde a_{1}(f,D) + O(\epsilon^{0}),
\ee
where
\begin{align}
\tilde{a}_0(f,D) & = \frac{1}{16\pi} \int dt d^2 x N \sqrt{h}\,f(t,x),\\
\tilde{a}_1(f,D) & = \frac{1}{48 \pi^{3/2}} \int dt d^{2}x N \sqrt{h}\,f(t,x) \left(R - \frac{1}{N}\Delta N \right).
\end{align}


\bibliographystyle{kp}
\bibliography{references}
\end{document}